# DYNAMICS OF THE BAR AT THE GALACTIC CENTRE

ORTWIN E. GERHARD

*Astronomisches Institut, Universität Basel*
*Venusstrasse 7, CH-4102 Binningen*

## 1.  Introduction

There is now substantial evidence for a rotating bar in the inner Galaxy. This is an important change in our perception of the Galaxy because it changes the ways in which we have to think about its evolutionary history.

The idea of a Galactic bar is not new; that motion on elliptic orbits in a barred potential might explain various aspects of the atomic and molecular gas observations near the Galactic Centre has been suggested a number of times (e.g., Peters 1975, Cohen & Few 1976, Liszt & Burton 1980, Gerhard & Vietri 1986, Mulder & Liem 1986, Sanders 1989).

What has changed in the past few years is (i) that the evidence now comes from several fronts, including the NIR light distribution as measured by COBE, IRAS source counts, atomic and molecular gas kinematics, the first indications for triaxiality in the stellar kinematics of the bulge, and perhaps the large optical depth to microlensing in the OGLE experiment; and (ii) that there is a dynamical model which provides a physical basis for explaining a number of independent features in the observed gas kinematics, is consistent to zeroth order with the other pieces of evidence for a bar that we now have, and promises to be extendable towards including these into one coherent picture.

In this review I give a brief summary of the current evidence for the bar in the inner Galaxy. Then I discuss in more detail the subject of gas flows in bars and in the Galactic Centre. The integrated NIR photometry is described in the paper by Dwek. Finally, several evolutionary processes are briefly discussed which become relevant for Galactic evolution because of the presence of the bar, such as gas infall, angular momentum transfer, subsequent central star formation, and the formation of peanut-bulge-like stellar systems from bars through a bending instability.

## 2. Evidence for the Galactic Centre Bar

Near-infrared photometry, IRAS source counts, and modelling of the atomic and molecular gas in the inner Galaxy all point to the existence of a $\sim 3:1$ elongated bar at the Galactic Centre with major axis length $2-4$ kpc and near side in the first quadrant of galactic longitude. Table 1 lists individual sources of evidence, references, length and axial ratios of the inferred bar, and the angle $\phi_{\mathrm{maj}}$ between the near side of the bar major axis and the line Galactic Centre - Sun, such that positive $\phi_{\mathrm{maj}}$ corresponds to position in the first quadrant. No information is listed when not constrained well or not known to this author. A short discussion of these observations follows.

The older balloon data (Matsumoto *et al.* 1982) and more clearly the COBE-DIRBE photometry (Weiland *et al.* 1994, Dwek *et al.* 1995) show that the Galactic bulge is both brighter and more extended in latitude at given positive longitude than at the same negative longitude, except for a region close to the Galactic Centre where the first effect is reversed. These signatures are just as expected for a triaxial bulge with its long axis in the first quadrant (Blitz & Spergel 1991), because then a line-of-sight at fixed positive $l$ cuts the the bar major axis at smaller galactocentric distance than one with the same negative longitude. The length given for the COBE bar is inferred from where the longitude asymmetry ends; it is not apparent in the Dwek *et al.* model fits. The apparent shape of the COBE bulge was matched by an N-body simulation of a peanut bulge formed from a disk-bar-bending instability (Sellwood 1993; see Section 4 below); this resulted in the parameters given in the third row. The vertical extent of the N-body bar is uncertain because of limited grid resolution.

Source counts in the bulge region come mostly from the IRAS data which is not restricted to a small number of special fields where the extinction is low. However, the very large number of survey stars in the OGLE experiment has allowed constraints on the spatial distribution to be obtained from just a few of these fields (Stanek *et al.* 1994). In the source samples listed in Table 1, the evidence for the bar comes from a distribution that is asymmetric in longitude with respect to the Sun-Galactic Centre line in either its number surface density or flux distribution, or both, as expected for an intrinsically barred distribution of objects observed at finite distance from the Sun. To determine the bar parameters accurately from such a sample requires sophisticated modelling (Weinberg 1992), taking into account the sample selection function, and large spatially extended samples.

Modelling the gas kinematics has the advantage of also constraining the pattern speed of the bar, $\Omega_p \simeq 60\,\mathrm{km}\,s^{-1}\,\mathrm{kpc}^{-1}$. These results are described in more detail in the next Section. The detailed comparison to observations depends on the choice of mass distribution for the bulge and bar (which is different for the two analyses listed). The BGSBU-model of Binney, Gerhard, Stark, Bally & Uchida (1991) is based on modelling the molecular paralellogram, HI terminal curve, and galactic centre cloud orbits, and uses a barred bulge with density $\propto r^{-1.8}$. The model of Weiner & Sellwood is described elsewhere in these proceedings.

| Evidence | from | Refs. | $\phi_{maj}$ | a (kpc) | a:b:c |
|----------|------|-------|--------------|---------|-------|
| NIR | Balloon | 1,2 | positive | > 1.8 | |
| | COBE DIRBE | 3,4 | $20° \pm 10°$ | 2.2 | $10:3 \pm 1:2 \pm 1$ |
| | N-body vs. COBE | 5 | $30°$ | $2.3^*$ | $10:3 - 4:2?$ |
| Discrete | IRAS Mira variables | 6 | $45°$ | $\sim 1.5$ | $4:1:1$ |
| Sources | IRAS sample | 7 | positive | | |
| | AGB stars | 8 | $36 \pm 10°$ | $\sim 4$ | $3:2:?$ |
| | Red clump stars | 9 | $< 45°$ | > 1 | |
| | Disk Glob. clusters | 10 | $25°$ | $\sim 3$ | $3:1:?$ |
| Gas | Parallelogram, HI | 11 | $16°$ | $2.2^*$ | |
| | HI-envelope | 12 | $\sim 30°$ | | |
| Kinematics | Bulge K-giants | 13 | | | |
| Microlensing | OGLE | 14-16 | $\lesssim 20$ | | $\gtrsim 3:1:?$ |

TABLE 1. Evidence for the bar in the inner galaxy. References are: 1. Matsumoto *et al.* 1982, 2. Blitz & Spergel 1991, 3. Weiland *et al.* 1994, 4. Dwek *et al.* 1995, 5. Sellwood 1993, 6. Whitelock & Catchpole 1992, 7. Nakada *et al.* 1991, 8. Weinberg 1992, 9. Stanek *et al.* 1994, 10. Blitz 1993, 11. Binney *et al.* 1991, 12. Weiner & Sellwood, these proceedings, 13. Zhao *et al.* 1994, 14. Paczynski *et al.* 1994a, 15. Evans 1994, 16. Zhao *et al.* 1995. $^*$: Estimated as $0.9\times$ corotation radius.

Indications for a barred bulge also come from the kinematics of bulge K-giant stars in Baade's window (Zhao, Spergel & Rich 1994), although the stellar sample is small and cannot be used to constrain the bar parameters quantitatively. The bar has been proposed as a key ingredient for explaining the high optical depth to gravitational lensing towards the bulge seen in the OGLE results (Paczynski *et al.* 1994a). In this case the bar must have its long axis near the Sun-Galactic Centre line. However, modelling results to-date are still controversial (Evans 1994, Zhao, Spergel & Rich 1995).

In summary, there now appears to be good evidence from several different sets of observations that our Galaxy contains an elongated ($\sim 3:1$) nuclear bar/bulge of length $2-4$ kpc, with its nearby long axis in the first longitude quadrant at $\phi_{maj} = 15 - 45°$. However, not all observations may measure the same physical bar population; e.g., the vertical scale-length of the Mira variables is significantly smaller than that of the COBE NIR emission. The bar parameters will undoubtedly be constrained much better when dynamical models based on the combined COBE NIR light distribution and HI and molecular gas dynamics have been constructed and have been used to predict source count results and microlensing probabilities.

Previously unknown central bars have recently been seen in a number of external galaxies with large-scale bars (*e.g.*, Wozniak *et al.* 1995, Shaw *et al.* 1995). The nuclear bar in the oval-disk galaxy M94 (Möllenhoff, Matthias & Gerhard 1995), e.g., has axis ratio in the galaxy plane $\sim 3 : 1$, ends at $\sim 1.2$ kpc somewhat inside an inner star-forming gas ring, and is less centrally concentrated and more flattened to the plane than the co-spatial axisymmetric part of the bulge. This example shows that also the structure of the Galactic bulge/bar could be considerably more complicated than the simple one-component triaxial models used sofar.

## 3. Gas Flow in the Inner Galaxy

Observations of cold gas in the inner galactic disk show a clumpy, asymmetric distribution, with large non-circular velocities up to $\sim 200 \,\mathrm{km\,s^{-1}}$. Only a small fraction ($\sim 10^7 \,\mathrm{M_\odot}$) of the cold gas in the Galactic Center (GC) region is in the form of neutral HI, measured at 21 cm (Burton & Liszt 1978, 1983; Sinha 1979; Braunsfurth & Rohlfs 1981). The dominant component ($\sim 10^8 \,\mathrm{M_\odot}$) is the cold molecular gas observed in mm-emission lines of molecules such as $^{12}$CO, $^{13}$CO, CS (Bania 1977, Sanders *et al.* 1984, Heiligman 1987, Dame *et al.* 1987, Bally *et al.* 1988). Despite the clumpiness and asymmetry of the gas distribution, the non-circular motions follow a coherent pattern whose gross properties can be described by assuming that the gas flows on tilted elliptical orbits (Peters 1975, Liszt & Burton 1980).

The *dynamical* understanding of such flow patterns comes from hydrodynamic simulations in gravitational fields. The key observation is that, if a quasi-equilibrium flow is established, it is generally a good approximation to think of such a flow in terms of the closed ballistic orbits in the underlying gravitational potential. This is independent of whether the simulations use sticky particles (Schwarz 1981, 1984; Habe & Ikeuchi 1985) or smooth fluids (Sanders & Huntley 1976; van Albada 1985; Mulder & Liem 1986; Athanassoula 1992). Physically, a cloud of gas released into a gravitational potential will shear and then settle onto closed orbits, because the energy in epicyclic motions is dissipated by collisions between fragment cloudlets. Subsequently, dissipation is reduced and the gas slowly drifts inwards along a sequence of closed orbits.

Only near resonances or where closed orbits intersect, do hydrody-namic forces significantly change this simple flow pattern. The location of the resonances in a barred potential is determined by the bar's pattern speed $\Omega_p$ and the circular velocity curve $v_c(R)$. These two parameters are therefore of primary importance for shaping the gas flow in a barred galaxy. For example, if the density and circular velocity curve scale as $\rho \propto R^{-\alpha}$ and $v_c(R) \propto R^{1-\alpha/2}$, then the corotation radius is proportional to $R_{CR} \propto \Omega_p^{-2/\alpha}$, and the ratios of the inner (ILR) and outer (OLR) Lindblad resonance radii to corotation is given by $R_{ILR}/R_{CR} = (1-\frac{1}{2}\sqrt{4-\alpha})^{2/\alpha}$ and $R_{OLR}/R_{CR} = (1+\frac{1}{2}\sqrt{4-\alpha})^{2/\alpha}$. Notice that these ratios are fairly sensitive to the parameter $\alpha$.

*Figure 1.* Closed orbits in a rotating barred potential and their projection into the $(l, v)$-diagram for an observer near the long axis of the bar. (From Gerhard 1992, after BGSBU).

Fig. 1 shows the sequence of prograde closed orbits inside corotation in a barred potential with circular velocity curve $v_c \propto R^{0.1}$, similar to that in the inner Galaxy. Gas inside corotation follows the main prograde '$x_1$' family until these orbits becomes self-intersecting near the inner Lindblad resonance. When the inflowing gas reaches the highest-energy self-intersecting $x_1$-orbit (the 'cusped orbit' in Fig. 1), the hydrodynamic simulations show that a shock forms, which causes it to switch to the closed '$x_2$' orbits deeper in the potential well. These are elongated along the potential's short axis.

The structure of such a shock near the ILR is shown in Fig. 2. The gas in this SPH-simulation is isothermal, and flows in a fixed rotating bar potential like that used by Athanassoula (1992). The velocity field outside the ILR region is clearly reminiscent of the closed orbit shapes in Fig. 1.

*Figure 2.* Morphology and velocity structure of gas flow in a rotating barred potential. Smooth Particle Hydrodynamic (SPH) simulation from Englmaier & Gerhard (in preparation).

Gas flowing inwards along the sequence of $x_1$-orbits eventually reaches the 'cusped' orbit. At pericentre on the 'cusped' orbit it crashes into gas at apocentre on the largest embedded $x_2$ orbit, producing a spray of material, which in turn causes a shock at the far side of the 'cusped' orbit. From there, material moves onto deeper-lying $x_2$ orbits, which are not affected by the shock. Across the shock in Fig. 2 the density jumps by a factor of $\simeq 5$, and the transverse velocity decreases by a similar factor. Van Albada (1985) and Athanassoula (1992) find similar velocity and density structures in their grid-based simulations. In fact, with enough resolution the results from the two different methods of solving the hydrodynamic equations are very similar.

BGSBU have argued that the key to interpreting the observations of cold gas in the inner Galaxy is the striking parallelogram in the $(l, v)$ plane seen in surveys of $^{12}$CO (Bania 1977) and $^{13}$CO (Heiligman 1987, Bally *et al.* 1988). BGSBU argue that this structure, shown in Fig. 3, is formed in a 'cusped' orbit shock like that in Fig. 2. In their model, the outermost $x_1$ orbits parallel to the long axis of the bar are mostly occupied by HI clouds. These clouds gradually drift inwards, onto more elongated orbits. Eventually they reach the cusped orbit, encounter the shock, in which most of the gas condenses into molecular form, and then plunge onto the small

*Figure 3.*   $(l, v)$ digram of $^{12}$CO $J = 1 \rightarrow 0$ averaged over $|b| < 0.1°$. The contours (spaced at intervals of 1 K) show a striking parallelogram. (From BGSBU.)

central $x_2$-orbits.

The projection of the sequence of closed $x_1$ and $x_2$ orbits into the $(l, v)$-diagram for an observer near the long axis of a rotating bar is shown in Fig. 1. Requiring that the $(l, v)$ trace of the 'cusped' orbit resemble the observed parallelogram determines that observers near the Sun must view the GC bar at a narrowly constrained angle around $\phi_{\mathrm{maj}} = 16°$. By contrast, only a weak limit on the bar's axial ratio $q$ can be obtained, as this does not influence the orbit shapes near the ILR much. With the same viewing angle, it is also possible to account for the rapid fall-off in the HI terminal velocity observed near $l = 2°$ and the subsequent slower decline out to $l = 12°$. For a given scale of the parallelogram and bulge mass profile, the HI data determine the pattern speed of the bar such that corotation is at $R_{\mathrm{CR}} \simeq 2.4 \pm 0.4$ kpc. In the region $R \lesssim 1.2$ kpc, where the HI terminal-velocity curve decreases outwards, the circular velocity is, in fact, rising. The inferred potential in the bulge region is given in Gerhard & Binney (1993).

The strong points of the BGSBU model are that it can fit the $(l, v)$ distribution and the associated large non-circular motions of the molecular parallelogram and the HI terminal curve. The locations of the giant molecular clouds at the GC, such as Sgr B and Sgr C, in the $(l, v)$-diagram are successfully predicted by the $x_2$-orbits in the model (Fig. 6 in Binney 1994). The model naturally accounts for the absence of cold gas between $R \sim 1.5$ kpc and $R \sim 3.5$ kpc, since in the neighbourhood of corotation no stable closed orbits exist on which cold gas could settle. The molecular ring at $R \sim 3.5$ kpc may be associated with gas accumulating near the bar's outer Lindblad resonance. Finally, the predicted perspective asymmetry in the gas distribution goes in the same sense as that observed, and in the sense expected from the bulge infrared photometry (Blitz & Spergel 1991, Weiland *et al.* 1994).

However, it also appears that some essential ingredients are still missing from the model. On the one hand, the observed lopsidedness of the GC molecular gas is probably too great to be accounted for just in terms of perspective effects. We are currently investigating whether asymmetries in the initial gas distribution can be enhanced by hydrodynamic clumping when the self-gravity of the gas is included. This problem is important in a wider context since many external barred galaxies show strong asymmetries in their gas flows. Another, probably harder, problem with the BGSBU model that it has no explanation for the tilt in the HI gas distribution (Liszt & Burton 1980). While simulations show that some gas in peanut-bulge potentials may settle on non-planar closed orbits (Friedli & Benz 1993, Friedli & Udry 1993), no obvious explanation of the HI tilt has yet emerged, mainly because the vertical parity of the orbits in these simulations is different from that of the tilt in the Galactic Centre.

A further uncertainty is the choice of 'correct' description for the interstellar gas in such simulations, as interstellar cloud hydrodynamics may not necessarily be very well described by the standard Euler equations (Binney & Gerhard 1993, Jenkins & Binney 1994). Jenkins & Binney have explored simple interaction schemes in sticky particle-hydrodynamic simulations of the Galactic gas flow. In these particle simulations, the interpretation of the molecular parallelogram in terms of the transition from $x_1$ to $x_2$ orbits is only partially born out, mainly, because no strong shocks form. This shows that the chosen description of interstellar gas does matter, and raises the interesting possibility of using observed gas flows to enhance our understanding of cloud fluids and *calibrate* the simulation schemes.

## 4. Evolutionary Processes in a Barred Milky Way

The presence of a bar in the inner Galaxy makes a number of evolutionary processes relevant for Galactic evolution. Some are briefly discussed in this Section.

### 4.1. GAS INFALL

In the rotating Galactic bar potential, molecular gas is moving inwards through the cusped orbit shock. Simple estimates for the present mass inflow rate through the ILR give $\sim 0.1\,\mathrm{M_\odot/\,yr}$ (Gerhard 1992). Further in, the massive GC clouds on $x_2$-orbits, such as Sgr B, circulate with speeds greater than the pattern velocity of the bar. Thus they tend to lose further angular momentum to the bulge/bar stars, both by gravitational torques from the rotating quadrupole, and by dynamical friction. The latter process alone causes gas inflow to yet smaller radii at a comparable rate (Stark *et al.* 1991). Thus the present population of Galactic centre clouds must be transient, and for the next Gyr material will be accreting onto the GC at an average rate of $\sim 0.01 - 0.1\,\mathrm{M_\odot/\,yr}$. Gas in the bulge region may be replenished by mass loss from the bulge stars ($\sim 0.2\,\mathrm{M_\odot/\,yr}$; Ciotti *et al.* 1991), and perhaps from gas moving inwards through corotation (if the medium is sufficiently dissipative to outway the spin-up from the bar there, or if the potential is time-varying; see, e.g., Schwarz 1981, Noguchi 1988, Pfenniger & Norman 1990, Friedli & Benz 1993). The frequent observation of central gas concentrations with associated non-circular motions in nearby barred spirals suggests that radial gas inflow is a common phenomenon (Kenney 1994).

### 4.2. ANGULAR MOMENTUM TRANSFER

The infalling gas must lose its angular momentum to the stars in the bar/bulge. For a hot stellar component the dynamical importance of this angular momentum transfer is enhanced by the fact that the typical circular velocity for gas in the bulge region is $\sim 180\,\mathrm{km\,s^{-1}}$, while the streaming velocity of the stars in, say, an oblate-isotropic bulge model is only $\sim 65\,\mathrm{km\,s^{-1}}$ (estimated from the model in Kent 1992). For a rapidly rotating stellar bar, the gas-driven angular momentum accretion may reverse the tendency of the bar to lose angular momentum to the surrounding spheroid or halo, which is predicted by dynamical friction calculations (Weinberg 1985, Hernquist & Weinberg 1992), as shown in the coupled gas- and stellardynamical simulations of Friedli & Benz (1993).

### 4.3. BUILD-UP OF THE INNER DISK

If accretion rates of $0.1\,\mathrm{M_\odot/\,yr}$ can be maintained over prolonged periods of time, and the material be mostly converted into stars, then significant additions to the mass budget of the inner disk will result. Over the age of the bulge $0.1\,\mathrm{M_\odot/\,yr}$ correspond to of order one third of the total mass inside 500 pc. The present star formation rate in the central few 100 pc is normal for the amount of gas there (as inferred from FIR and thermal radio emission, Güsten 1989). It may currently be suppressed by the pressure from a very hot, $10^8$K gas (Spergel & Blitz 1992, Blitz *et al.* 1993), but it is hard to see how this hot medium can be maintained over long time-scales without converting a significant fraction of the accreted material into stars, unless unusually many massive stars are produced. Kenney (1994) discusses the triggering of burst-like star formation by gas inflow in external barred galaxies.

## 4.4.  BAR DISSOLUTION BY A CENTRAL MASS CONCENTRATION

By funneling material to its centre, the bar may generate the cause of its own destruction: Hasan & Norman (1990) and Hasan, Pfenniger & Norman (1993) have shown that a central mass with a few percent of the bar mass ($\sim 10^8 \, \mathrm{M}_\odot$ in our Galaxy) would destroy the bar. For a mass of this order the outer ILR has moved outwards sufficiently, so that there no longer exist the elongated $x_1$-orbits to self-consistently support the bar's elongated shape. While the Galaxy's central point mass is not (yet) relevant for such evolution of the bulge/bar, Friedli & Benz (1993) and Friedli & Martinet (1993) have shown that bar-driven gas inflow may form sufficiently concentrated gas disks or secondary bars which can have similar effects.

## 4.5.  LATE BULGE FORMATION FROM DISK INSTABILITIES

It has recently emerged from N-body simulations that two-dimensional bars are often unstable to a bending instability and evolve to form peanut-shaped bulges (Combes *et al.* 1990, Raha *et al.* 1991, Pfenniger & Friedli 1991; for a review see Sellwood & Wilkinson 1993). Pfenniger & Friedli (1991) show that the final peanut-shaped bulge is largely supported by a 2:2:1 resonant orbital family antisymmetric with respect to the equatorial plane. The instability appears to be wide-spread; most strong bars are quickly become three-dimensional. The mechanism appears to work for thin and thick initial disk bars, and for a range of initial disk-to-preexisting bulge ratios. After their formation, these peanut-bulges apparently do not evolve significantly (Pfenniger & Friedli 1991).

The resulting peanut bulges are centrally concentrated, substantially flattened bodies. Seen edge-on, they show clear peanut-shapes, but have more normal, nearly elliptical shapes when viewed end-on, and boxy shapes at intermediate angles (Combes *et al.* 1990). Their dynamical structure is closer to the dynamics of disks than classical spheroidal bulges; it is governed by fast rotation nearly constant on cylinders. This is similar as in some observed peanut bulges such as in NGC 4565 (Kormendy & Illingworth 1982). Sellwood's (1993) model has a remarkably similar morphology to the COBE bulge/bar, and a kinematic anisotropy that is not too dissimilar to that deduced from the observed radial velocities (summarized by Kent 1992) and proper motions in Baade's window (Spaenhauer, Jones & Whitford 1992).

One may speculate that, integrated over the age of the Galaxy, a significant part of the Galactic bulge/bar might have been made in the disk and scattered out of the galactic plane in this way. The test of this hypothesis will come from observations of ages, abundances, and kinematics of bulge stars, a difficult subject reviewed by Rich (1993) and other papers in Dejonghe & Habing (1993). While the evidence points towards stellar ages of $\sim 5 - 10$ Gyr in Baade's window, with an age spread (e.g., Holtzman *et al.* 1993, Paczynski *et al.* 1994b), the situation at lower latitudes is less clear. The correlations between kinematics and [Fe/H] argue for dissipative, perhaps extended formation of the bulge (Rich 1990, Minniti 1995). Clearly, much remains to be learned here.

# References

1.  Athanassoula E., 1992, MNRAS **259**, 345
2.  Bania T.M., 1977. ApJ **216**, 381
3.  Bally J., Stark A.A., Wilson R.W., Henkel C., 1988, ApJ **324**, 223
4.  Binney J.J., 1994, in The Nuclei of Normal Galaxies, NATO ASI Ser. C, Vol. 445, Genzel R., Harris A.I., eds, Kluwer, Dordrecht, 75
5.  Binney J.J., Gerhard O.E., 1993, in Back to the Galaxy, AIP Conf. Proc. 278, Holt S.S., Verter F., eds, AIP, New York, 87
6.  Binney J.J., Gerhard O.E., Stark A.A., Bally J., Uchida K.I., 1991, MNRAS **252**, 210 (BGSBU)
7.  Blitz L., 1993, in Back to the Galaxy, AIP Conf. Proc. 278, Holt S.S., Verter F., eds, AIP, New York, 98
8.  Blitz L., Binney J., Lo K.Y., Bally J., Ho P.T.P., 1993, Nature **361**, 417
9.  Blitz L., Spergel D.N., 1991, ApJ **379**, 631
10. Braunsfurth E., Rohlfs K., 1981, A&AS **44**, 437
11. Burton W.B., Liszt H.S., 1978, ApJ **225**, 815
12. Burton W.B., Liszt H.S., 1983, A&AS **52**, 63
13. Ciotti L., D'Ercole A., Pellegrini S., Renzini A., 1991, ApJ **376**, 380
14. Cohen R.J., Few R.W., 1976, MNRAS **176**, 495
15. Combes F., Debbasch F., Friedli D., Pfenniger D., 1990, A&A **233**, 82
16. Dame T.M. etal, 1987, ApJ **322**, 706
17. Dejonghe H., Habing H.J., eds, 1993, Galactic Bulges, Proc. IAU Symp. 153, Kluwer, Dordrecht
18. Dwek E. *et al.*, 1995, ApJ, in press
19. Evans N.W., 1994, ApJL **437**, L31
20. Friedli D., Benz, W, 1993, A&A **268**, 65
21. Friedli D., Martinet L., 1993, A&A **277**, 27
22. Friedli D., Udry, S., 1993, in Galactic Bulges, Proc. IAU Symp. 153, Dejonghe H., Habing H.J., eds, Kluwer, Dordrecht, 273
23. Gerhard O.E., 1992, Reviews in Modern Astronomy **5**, 174
24. Gerhard O.E., Binney J.J., 1993, in Galactic Bulges, Proc. IAU Symp. 153, Dejonghe H., Habing H.J., eds, Kluwer, Dordrecht, 275
25. Gerhard O.E., Vietri M., 1986, MNRAS **223**, 377

26.  Güsten R., 1989, in The Center of the Galaxy, Proc. IAU Symp. 136, Morris M.,
     ed, Kluwer, Dordrecht, 89
27.  Habe A., Ikeuchi S., 1985, ApJ **289**, 540
28.  Hasan H., Norman C., 1990, ApJ **361**, 69
29.  Hasan H., Pfenniger D., Norman C., 1993, ApJ **409**, 91
30.  Heiligman G.M., 1987, ApJ **314**, 747
31.  Hernquist L., Weinberg M.D., 1992, ApJ **400**, 80
32.  Holtzman J.A. *et al.* , 1993, AJ **106**, 1826
33.  Jenkins A., Binney J.J., 1994, MNRAS **270**, 703
34.  Kenney J., 1994, in Astronomy with Millimeter and Submillimeter Wave Interfer-
     ometry, Proc. IAU Coll. 140, Ishiguro M., ed, ASP Conf. Series, ASP, San Francisco
35.  Kent S.M., 1992, ApJ **387**, 181
36.  Kormendy J., Illingworth G., 1982, ApJ **256**, 460
37.  Liszt H.S., Burton W.B., 1980, ApJ **236**, 779
38.  Matsumoto T. *et al.* , 1982, in The Galactic Center, AIP Conf. Proc. 83, Riegler
     G.R., Blandford R.D., eds, AIP, New York, 48
39.  Minniti D., 1995, ApJ , submitted
40.  Möllenhoff C., Matthias M., Gerhard O.E., 1995, A&A , in press
41.  Mulder W.A., Liem B.T., 1986. A&A **157**, 148
42.  Nakada Y., Deguchi S., Hashimoto O., Izumiura H., Onaka T., Seki- guchi K.,
     Yamamura I., 1991, Nature **353**, 140
43.  Noguchi M., 1988, A&A **203**, 259
44.  Paczynski B., Stanek K.Z., Udalski A., Szymanski M., Kaluzny J., Kubiak M.,
     Mateo M., Krzeminski W., 1994a, ApJL **435**, L113
45.  Paczynski B., Stanek K.Z., Udalski A., Szymanski M., Kaluzny J., Kubiak M.,
     Mateo M., 1994b, AJ **107**, 2060
46.  Peters W.L., 1975, ApJ **195**, 617
47.  Pfenniger D., Norman C., 1990, ApJ **363**, 391
48.  Pfenniger, D., Friedli, D., 1991, A&A **252**, 75
49.  Raha N., Sellwood J.A., James R.A., Kahn F.D., 1991. Nature **352**, 411
50.  Rich R.M., 1990, ApJ **362**, 604
51.  Rich R.M., 1993, in Galactic Bulges, Proc. IAU Symp. 153, Dejonghe H., Habing
     H.J., eds, Kluwer, Dordrecht, 169
52.  Sanders D.B., Solomon P.M., Scoville N.Z., 1984, ApJ **276**, 182
53.  Sanders R.H., 1989, in The Center of the Galaxy, Proc. IAU Symp. 136, Morris M.,
     ed, Kluwer, Dordrecht, 77
54.  Sanders R.H., Huntley J.M., 1976, ApJ **209**, 53
55.  Schwarz M.P., 1981, ApJ **247**, 77
56.  Schwarz M.P., 1984, MNRAS **209**, 93
57.  Sellwood J.A., 1993, in Back to the Galaxy, AIP Conf. Proc. 278, Holt S.S., Verter
     F., eds, AIP, New York, 133
58.  Sellwood J.A., Wilkinson A., 1993, RepProgPhys **56**, 173
59.  Shaw M., Axon D., Probst A., Gatley I., 1995, MNRAS , in press
60.  Sinha R.P., 1979, A&AS **37**, 403
61.  Spaenhauer A., Jones B.F., Whitford A.E., 1992, AJ **103**, 297
62.  Spergel D.N., Blitz L., 1992, Nature **357**, 665
63.  Stanek K.Z., Mateo M., Udalski A., Szymanski M., Kaluzny J., Kubiak M., 1994,
     ApJL **429**, L73
64.  Stark A.A., Gerhard O.E., Binney J.J., Bally J., 1991. MNRAS **248**, 14P
65.  van Albada G.D., 1985, A&A **142**, 491
66.  Weiland J.L. *et al.* , 1994, ApJL **425**, L81
67.  Weinberg M.D., 1985, MNRAS **213**, 451
68.  Weinberg M.D., 1992, ApJ **384**, 81
69.  Whitelock P., Catchpole R., 1992, in The Center, Bulge, and Disk of the Milky Way,
     Blitz L., ed, Kluwer, Dordrecht, 103

70.   Wozniak H., Friedli D., Martinet L., Martin P., Bratschi P., 1995, A&A , in press
71.   Zhao H.-S., Spergel D.N., Rich R.M., 1994, AJ **108**, 2154
72.   Zhao H.-S., Spergel D.N., Rich R.M., 1995, ApJ , submitted